\documentclass[11pt,twocolumn, notitlepage]{article}
\usepackage[utf8]{inputenc}
\usepackage{soul}
\usepackage{authblk}
\usepackage{cite}
\usepackage{url}
\usepackage{xcolor}
\usepackage{graphicx}
\usepackage{amsmath}
\usepackage{amssymb}
\usepackage{wasysym}
\usepackage{bm}
\usepackage{adjustbox}
\usepackage[T1]{fontenc}
\usepackage{babel}
\usepackage{siunitx}
\usepackage{pgfplots}
\usepgfplotslibrary{groupplots}
\pgfplotsset{compat=newest}
\pgfplotsset{
  every axis plot/.append style={very thick},
}
\pgfplotsset{width=10cm}
\usepackage[margin=1in]{geometry}
\graphicspath{./images/}

\title{Interaction between pentacene molecules and monolayer transition metal dichalcogenides}
\author[1]{E. Black}
\author[2]{P. Kratzer}
\author[1, *]{J. M. Morbec}
\affil[1]{School of Chemical and Physical Sciences, Keele University, Keele ST5 5BG, UK}
\affil[2]{Fakult\"at f\"ur Physik, Universit\"at Duisburg-Essen, Campus Duisburg, Lotharstr. 1, 47057 Duisburg, Germany}
\affil[*]{Corresponding author: j.morbec@keele.ac.uk}
\date{}
\begin{document}
\twocolumn[
\begin{@twocolumnfalse}
\maketitle
\begin{abstract}
	Using first-principles calculations based on density-functional theory, we investigated the adsorption of pentacene molecules on monolayer two-dimensional transition metal dichalcogenides (TMD). 
	We considered the four most popular TMDs, namely, MoS$_2$, MoSe$_2$, WS$_2$ and WSe$_2$, and we examined the structural and electronic properties of pentacene/TMD systems. We discuss how monolayer pentacene interacts with the TMDs, and how this interaction affects the charge transfer and work function of the heterostructure. We also analyse the type of band alignment formed in the heterostructure and how it is affected by molecule-molecule and molecule-substrate interactions. Such analysis is valuable since pentacene/TMD heterostructures are considered to be promising for application in flexible, thin and lightweight photovoltaics and photodetectors. 
    \end{abstract}
    \vspace{1cm}
\end{@twocolumnfalse}
]

\section{Introduction}
Monolayer two-dimensional transition metal dichalcogenides (TMD) such as MoS$_2$, MoSe$_2$, WS$_2$ and WSe$_2$ have emerged in the past decade as promising materials for a variety of applications, ranging from photovoltaics\cite{Wi-ACSNano-2014} and photodetectors\cite{Du-nanotechnology-2020} to gas sensors\cite{Lee-ACSSens-2018} and DNA sequencing\cite{Perez-PCCP-2020}. In bulk crystals, TMDs form layered structures where the layers are held together via van der Waals (vdW) forces. When exfoliated or synthesized in two-dimensional (2D) form, the monolayers have no dangling bonds, making it easy to combine them with other systems forming vdW heterostructures. Due to the non-directional nature of vdW forces, 2D materials can be combined with materials of different dimensionalities such as quantum dots\cite{Tanoh-ACSNano-2020} and nanotubes\cite{Liu-Nanoscale-2019}, which may result in heterostructures with better properties and functionalities than the individual components. 

Organic molecules are particularly interesting materials to combine with 2D systems. The large library of known molecules, which includes donors and acceptors as well as excellent absorbers and photo- and thermo-responsive molecules, 
offers a wide variety of systems that can be employed to enhance the properties and modify the functionalities of 2D materials; 
for instance, the adsorption of F$_4$TCNQ and PTCDA molecules has been found to turn monolayer MoS$_2$ into a p-type semiconductor \cite{Le-RSCAdv-2021} and to enhance its photoluminescence intensity \cite{Mouri-NanoLett-2013, Habib-Nanoscale-2018}. 
Additionally, both 2D and organic materials are flexible, thin and lightweight systems, which makes organic/2D heterostructures especially attractive for wearable and portable applications. 

Pentacene (PEN) is one of the most popular organic materials, largely investigated for optoelectronic and photovoltaic applications due to its high carrier mobility\cite{Nelson-APl-1998}, intense photoluminescence\cite{He-APL-2010}, excellent photosensitivity\cite{Kim-APL-2003} and strong absorption in the visible range of the solar spectrum\cite{Maliakal-ChemMat-2004}. It has been recently reported that pentacene and MoS$_2$ form p-n type-II heterojunction, with ultrafast charge transfer and long-lived charge-separated state\cite{Homan-NanoLett-2017}. As revealed by Homan et al.\cite{Homan-NanoLett-2017}, pentacene/MoS$_2$ heterostructure exciton dissociation occurs by  hole  transfer  to  pentacene  on  the  time  scale  of 6.7~ps,  fast  enough  to  surpass  most  of  the  hole  relaxation  processes  and  yield  a  net  hole transfer of ~50\% in the heterojunction, and the charge-separated state lives for approximately 5.1~ns, 2–60 times longer than the recombination lifetimes previously reported for 2D/2D vdW heterostructures such as MoSe$_2$/WS$_2$, MoS$_2$/MoSe$_2$ and MoSe$_2$/WSe$_2$\cite{Homan-NanoLett-2017}. 
This finding suggests that pentacene/MoS$_2$ heterostructures (and potentially pentacene/MoSe$_2$, pentacene/WS$_2$ and pentacene/WSe$_2$ heterostructures) are promising for optoelectronic and photovoltaic applications.  

Despite the great potential of pentacene/TMD heterostructures for technological applications, a systematic investigation of the interaction between monolayer TMDs and pentacene molecules, in particular in the monolayer regime, has not yet been performed. Existing studies either focus on pentacene molecules adsorbed only on MoS$_2$\cite{Shen-AdvMaterInterfaces-2017, Kachel-ChemSci-2021} or on pentacene films adsorbed on TMDs\cite{Habib-AdvTheorySimul-2020}, notwithstanding a recent experimental work\cite{Kachel-ChemSci-2021} reporting that monolayer pentacene on MoS$_2$ is thermally stabilized compared to multilayer pentacene.  
This work, therefore, aims to fill this gap by presenting a first-principles study, based on density-functional-theory calculations, of the adsorption of single-layer pentacene molecules on monolayer 2D TMDs (MoS$_2$, MoSe$_2$, WS$_2$ and WSe$_2$). 
We considered one monolayer molecular coverage and we examined the structural and electronic properties of pentacene/TMD heterostructures. 
We examined the interaction between monolayer pentacene and TMD, and how this interaction affects the charge transfer, work function and band alignment of the pentacene/TMD heterostructures.

\section{Computational Details}

Density-functional-theory (DFT)~\cite{Kohn-PR140-A1133} calculations were performed using the Quantum ESPRESSO suite\cite{giannozzi2009quantum, giannozzi2017advanced}, which employs plane-wave basis sets and pseudopotentials. The generalized gradient approximation (GGA) proposed by Perdew, Burke, and Ernzerhof (PBE)\cite{perdew1996generalized} was used for the exchange-correlation functional alongside Grimme's DFT-D3 van der Waals force corrections \cite{grimme2010consistent}. 
Projector augmented wave (PAW)\cite{Blochl-PRB-1994} pseudopotentials\cite{PSlibrary}  
were used with wavefunction energy cutoffs (set after energy convergence tests) of 80, 100, 80 and 120~Ry for MoS$_2$, MoSe$_2$, WS$_2$ and WSe$_2$, respectively. The pentacene/TMD heterostructures were modeled within the supercell approach, considering a $7\times 4$ cell for the TMD and a vacuum region of about 40~\AA \ along the direction perpendicular to the monolayer TMD plane. Truncation of the Coulomb interaction in this direction, as proposed by Sohier et al.\cite{Sohier-PRB-2017} for 2D systems, was employed in all calculations. The Brillouin zone was sampled using a $3\times 6\times 1$ Monkhorst-Pack\cite{Monkhorst-PRB1976} k-point grid, determined after energy convergence tests.  
For the lattice parameters of the TMD systems, we used the calculated values of 3.17, 3.30, 3.17 and 3.29~\AA \ for MoS$_2$, MoSe$_2$, WS$_2$ and WSe$_2$, respectively, 
which are in good agreement with DFT values reported in the literature for bulk (namely, 3.16, 3.29, 3.16 and 3.28~\AA, respectively)\cite{Bastos-PRM-2019} and 2D monolayer (namely, 3.183, 3.318, 3.182, 3.315~\AA, respectively)\cite{Duerloo-NatCommun-2014} systems. A summary of the computational details can be found in Table S1 of Supplementary Material.  

\section{Results}

Adsorption of pentacene on monolayer 2D TMD has been investigated considering one pentacene molecule in a $7\times 4$ supercell, which results in a minimum molecule-molecule distance (for horizontally oriented molecules) of approximately 6.2~\AA \ for MoS$_2$- and WS$_2$-based systems and 6.5~\AA \ for MoSe$_2$- and WSe$_2$-based systems. 
We have initially considered both horizontally and vertically oriented adsorption, and we found that vertical adsorption (both with long and short axes of pentacene molecule oriented parallel to the TMD plane) is highly unfavourable, with total energies at least 0.75~eV higher than the most favourable horizontal configuration. This agrees with recent experimental\cite{Kachel-ChemSci-2021} and theoretical\cite{Shen-AdvMaterInterfaces-2017} work reporting that pentacene molecules in the monolayer regime lie flat on the MoS$_2$ surface; in particular, the theoretical work based on DFT calculations for pentacene/MoS$_2$ revealed that vertical adsorption (with the long axis of pentacene lying parallel to MoS$_2$) is 0.6~eV higher in energy than the horizontal orientation. 
Therefore, herein we will only focus on horizontal adsorption. 

\begin{figure*}[!h]
	\centering
	\includegraphics[scale=1.0]{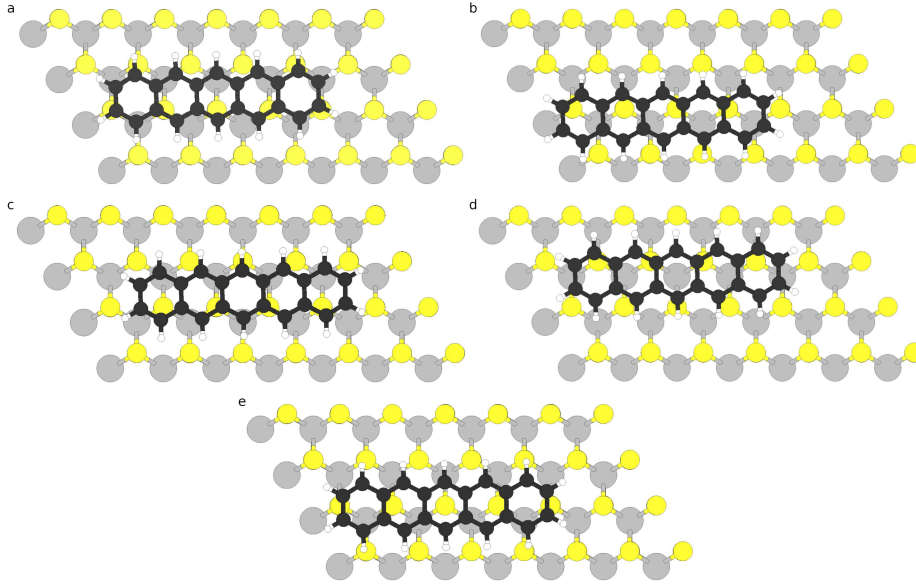}
	\caption{\label{fig-adsorptionsites}Ball and stick representation of the adsorption sites examined for the adsorption of pentacene molecule (horizontally oriented) on 2D monolayer TMD: (a) bridge-A, (b) bridge-B, (c) hollow, (d) top-TM, and (e) top-Ch. The adsorption sites were named based on the position of the central ring of the pentacene molecule.}
\end{figure*}

We have examined five adsorption sites, as shown in Fig.\ref{fig-adsorptionsites}, based on the position of the central ring of the pentacene molecule: two bridge sites: bridge-A (Fig.\ref{fig-adsorptionsites}(a)) and bridge-B (Fig.\ref{fig-adsorptionsites}(b)), where the central ring of pentacene lies over a bond between the transition metal (Mo or W) and the chalcogen atoms (S or Se); hollow site (Fig.\ref{fig-adsorptionsites}(c)), where the central ring of pentacene is on top of a hexagon in the TMD cell; top-TM (Fig.\ref{fig-adsorptionsites}(d)), where the central ring of pentacene is on top of a transition metal (Mo or W) atom; and top-Ch (Fig.\ref{fig-adsorptionsites}(e)), where the central ring of pentacene is on top of chalcogen (S or Se) atom. For the bridge sites,  we considered two configurations: bridge-A (Fig.\ref{fig-adsorptionsites}(a)) and bridge-B (Fig.\ref{fig-adsorptionsites}(b)), which resulted in the atoms of the molecule being located in different sites on the TMD. 
After geometry optimization (in which the internal coordinates of both molecule and TMD were allowed to relax), top-Ch was found to be the most favourable adsorption site for all of the TMDs, followed very close by the bridge B configuration, which is less than 6~meV higher in energy (see Table S2 in the Supplementary Material). The reason why top-Ch and bridge-B adsorption sites are more favourable may be due to the fact that in such configurations there are more C atoms from pentacene sitting on hollow sites of the TMD and less C atoms sitting on top of S/Se sites, which reduces the steric repulsion between pentacene C atoms and TMD chalcogen atoms. 
The other adsorption sites are between 24 and 83~meV higher in energy than top-Ch configuration (see Table S2 in the Supplementary Material). 
This small difference in energy among the different adsorption sites indicate that the molecules may be highly mobile in the single layer regime, as has also been suggested in Ref.~\cite{Kachel-ChemSci-2021}. 
The pentacene molecule was found to lie flat in all four TMDs, without any significant tilting or bending. 
The distances between the centre of mass of the pentacene molecule in top-Ch configuration and the top chalcogen layer are in the range of 3.3 and 3.4~\AA, as listed in Table~\ref{table-adsorption-energies-distances}, which is in good agreement with the result obtained via DFT calculations (3.4~\AA) for pentacene/MoS$_2$\cite{Shen-AdvMaterInterfaces-2017}. We note that the distances for Se-systems (MoSe$_2$ and WSe$_2$) are 0.09~\AA \ larger than those for S-systems, even though the adsorption energies are also larger for those systems (as will be discussed in the next paragraph). This difference, which has also been observed in other organic/TMD heterostructures~\cite{Krumland-ElectronStruc-2021}, may be due to the larger vdW radius of Se atom (1.90~\AA\cite{Batsanov-InorgMat-2001}) when compared to S atoms (1.73~\AA\cite{Batsanov-InorgMat-2001}). We also observe that the less favourable adsorption sites (namely, hollow, bridge-B, and top-TM) have larger adsorption distances (between 0.06 and 0.08~\AA as listed in Table~S3 of the Supplementary Material) than those found for top-Ch configuration; this is also a result of the steric repulsion between the C atoms of pentacene and the chalcogen atoms of the TMD, which causes larger adsorption distances and is expected to be more significant in hollow, bridge-A and top-TM configurations where several pentacene C atoms are located on top of the TMD chalcogen atoms.

Adsorption energies ($E_{\rm ads}$) of pentacene/TMD systems were calculated as the difference between the total energy of the combined system ($E_{\rm PEN/TMD}$) and the total energies of the isolated systems ($E_{\rm TMD}^{\rm relax}$ and $E_{\rm PEN}^{\rm iso-relax}$) in their relaxed geometries: 
\begin{equation}
	E_{\rm ads} = E_{\rm PEN/TMD} - E_{\rm TMD}^{\rm relax} - E_{\rm PEN}^{\rm iso-relax}.
\label{eq-ads-energies}
\end{equation}
$E_{\rm TMD}^{\rm relax}$ was computed considering the TMD systems in a $7\times 4$ supercell with their geometries optimized in the absence of the pentacene molecules.  
$E_{\rm PEN}^{\rm iso-relax}$ was obtained considering one single pentacene molecule in a cubic supercell with lateral dimension of 48~\AA \  and allowing all the atomic positions to relax. 
This way of calculating the adsorption energy ensures that $E_{\rm ads}$ includes contributions from molecule-molecule interactions, deformation of the TMD systems, deformation of the pentacene molecules, in addition to molecule-substrate interaction. 
A negative value for $E_{\rm ads}$ indicates that the formation of hybrid pentacene/TMD is energetically favourable.  As can be seen in Table~\ref{table-adsorption-energies-distances}, single-layer pentacene molecules are expected to favourably adsorb on all four TMDs, with adsorption energies between 1.38 and 1.46~eV for top-Ch adsorption sites, which is slightly smaller than the value (1.6~eV) calculated in Ref.\cite{Shen-AdvMaterInterfaces-2017} for pentacene/MoS$_2$. We note that pentacene molecules bind more strongly with Se-systems when compared with the S-counterparts, and with W-systems when compared with Mo-systems. Further analysis on the electronic properties and charge density of the PEN/TMD heterostructures will try to shine a light on the reason for that.

\begin{table*}[!h]
	\caption{\label{table-adsorption-energies-distances}Adsorption energies ($E_{\rm ads}$) and adsorption distances ($d$) of pentacene/TMD heterostructures (with pentacene adsorbed on top-Ch configuration) obtained with (PBE+vdW) and without (PBE only) including vdW correction methods.  
$E_{\rm ads}$ was computed using in Eq.~\ref{eq-ads-energies}. $d$ was obtained as the distance between the centre of masses of the pentacene molecule and the top chalcogen layer of the TMD systems.}
\centering
\begin{tabular}{lccccc}
\hline
	& \multicolumn{2}{c}{PBE+vdW} && \multicolumn{2}{c}{PBE only}\\
\cline{2-3} \cline{5-6}
	system       &  $E_{\rm ads}$ (eV) & $d$ (\AA) && $E_{\rm ads}$ (eV) & $d$ (\AA) \\
\hline
	PEN/MoS$_2$ & -1.389 & 3.309 && -0.100 & 3.988 \\
	PEN/MoSe$_2$ & - 1.425 & 3.400 && -0.101 & 4.020\\
	PEN/WS$_2$ & -1.435 & 3.297 && -0.097 & 4.049\\
	PEN/WSe$_2$ & -1.458 & 3.378 && -0.099 & 4.174 \\
\hline
\end{tabular}
\end{table*}

From the results displayed in Table~\ref{table-adsorption-energies-distances} we also observe that most of the adsorption energies are due to vdW interactions; when vdW interactions are switched off (see ``PBE only'' results in Table~\ref{table-adsorption-energies-distances}), the adsorption energies are reduced by one order of magnitude and the differences among the different TMD systems become even smaller. 
We also observe that vdW interactions bring the pentacene molecules closer to the substrate: PBE-only adsorption distances are at least 0.6~\AA \ larger than those obtained including vdW corrections. 

We can also examine the contributions to the adsorption energy from molecule-molecule and molecule-substrate interactions as well as from deformation of molecule and substrate (see Supplementary Materials for details of these calculations). As shown in Table~\ref{table-contributions}, the largest contribution to the adsorption energy comes from the molecule-substrate interaction in contrast with the negligible contribution from molecule-molecule interaction. We also observe small values for the energy associated with the deformation of the molecules and the substrates. In fact, we did not observe any bending of the molecule under the adsorption process and the C-C bond length in pentacene changes by less than 0.1\% in all four heterostructures when compared with isolated pentacene. For the TMDs, we observed small contractions ($<0.1$\%) in the bond lengths between transition metal and chalcogen atoms, mostly in the region where the pentacene is adsorbed.

\begin{table*}[!h]
	\caption{\label{table-contributions}Contributions to the adsorption energy from molecule-molecule interaction, molecule-substrate interaction, deformation of the molecule and deformation of the substrate. Energies are given in eV.}
\centering
\begin{tabular}{lcccc}
\hline
	       &  PEN/MoS$_2$ & PEN-MoSe$_2$ & PEN-WS$_2$ & PEN-WSe$_2$ \\
\hline
	molecule-molecule interaction & -0.004 & -0.003 & -0.004 & -0.003 \\
	molecule-substrate interaction & -1.402 & -1.423 & -1.340 & -1.452 \\
	molecule-deformation & 0.005 & -0.007 & -0.003 & -0.007 \\
	substrate-deformation & 0.012 & 0.008 & -0.088 & 0.004 \\
\hline
\end{tabular}
\end{table*}

\begin{figure*}[!h]
	\centering
	\includegraphics[scale=0.3]{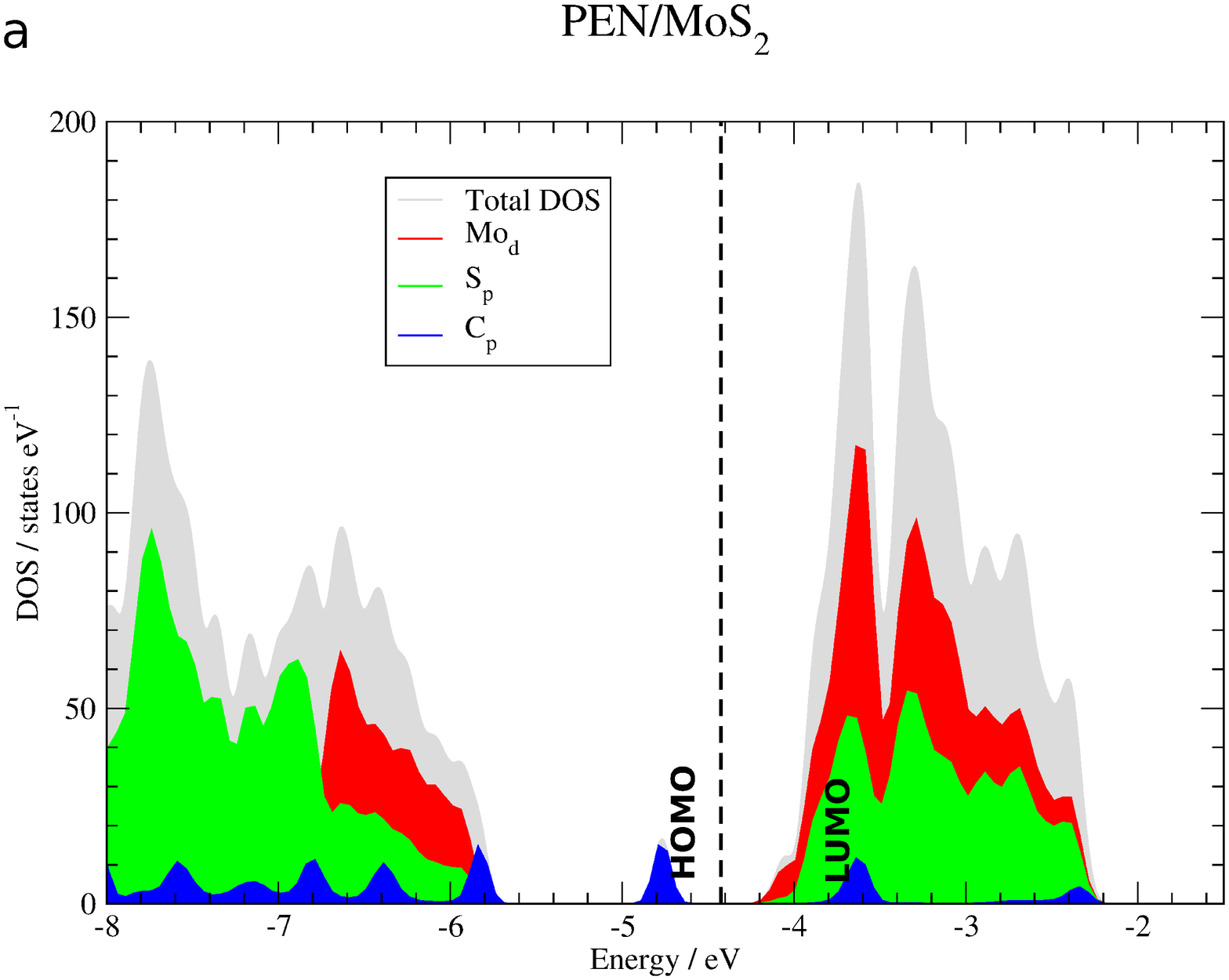}
	\includegraphics[scale=0.6]{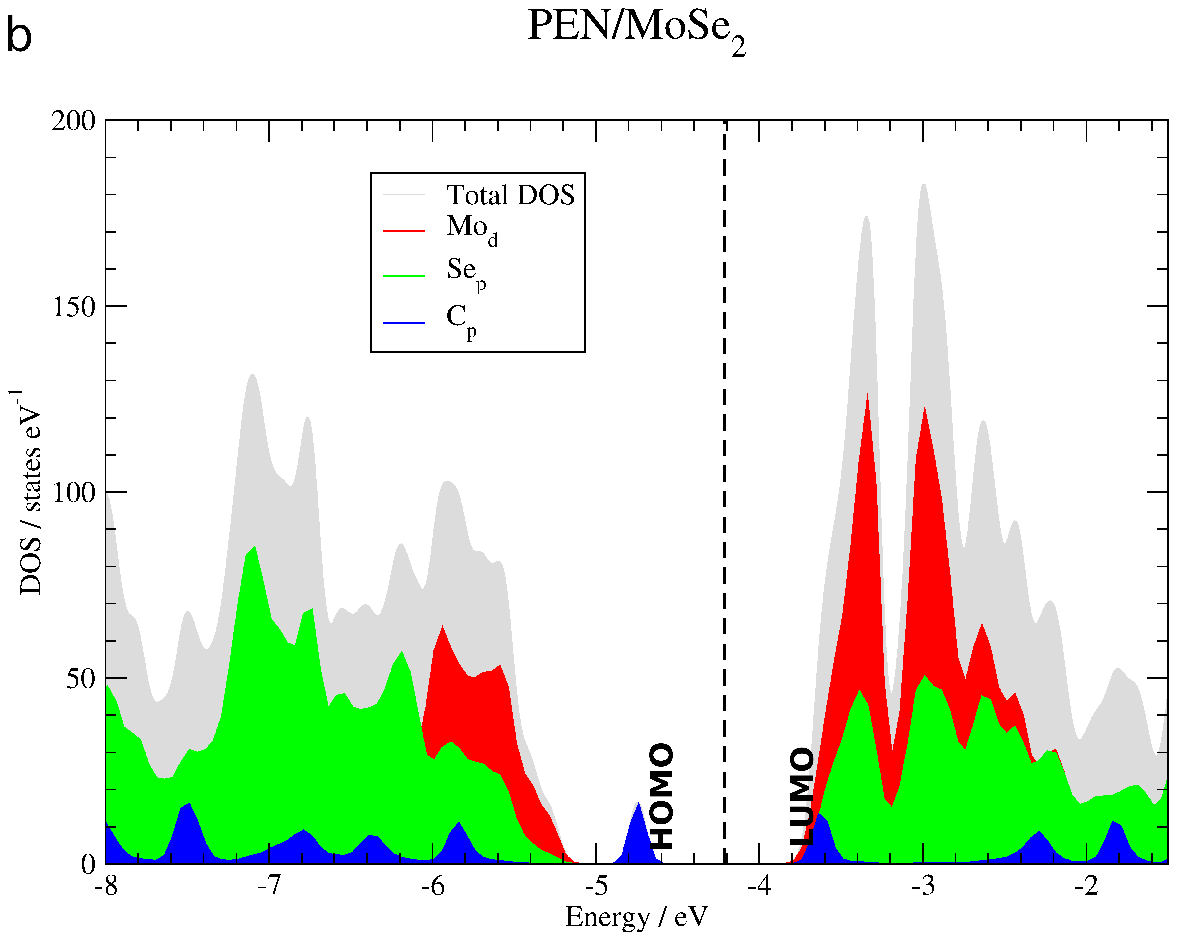}
	\includegraphics[scale=0.6]{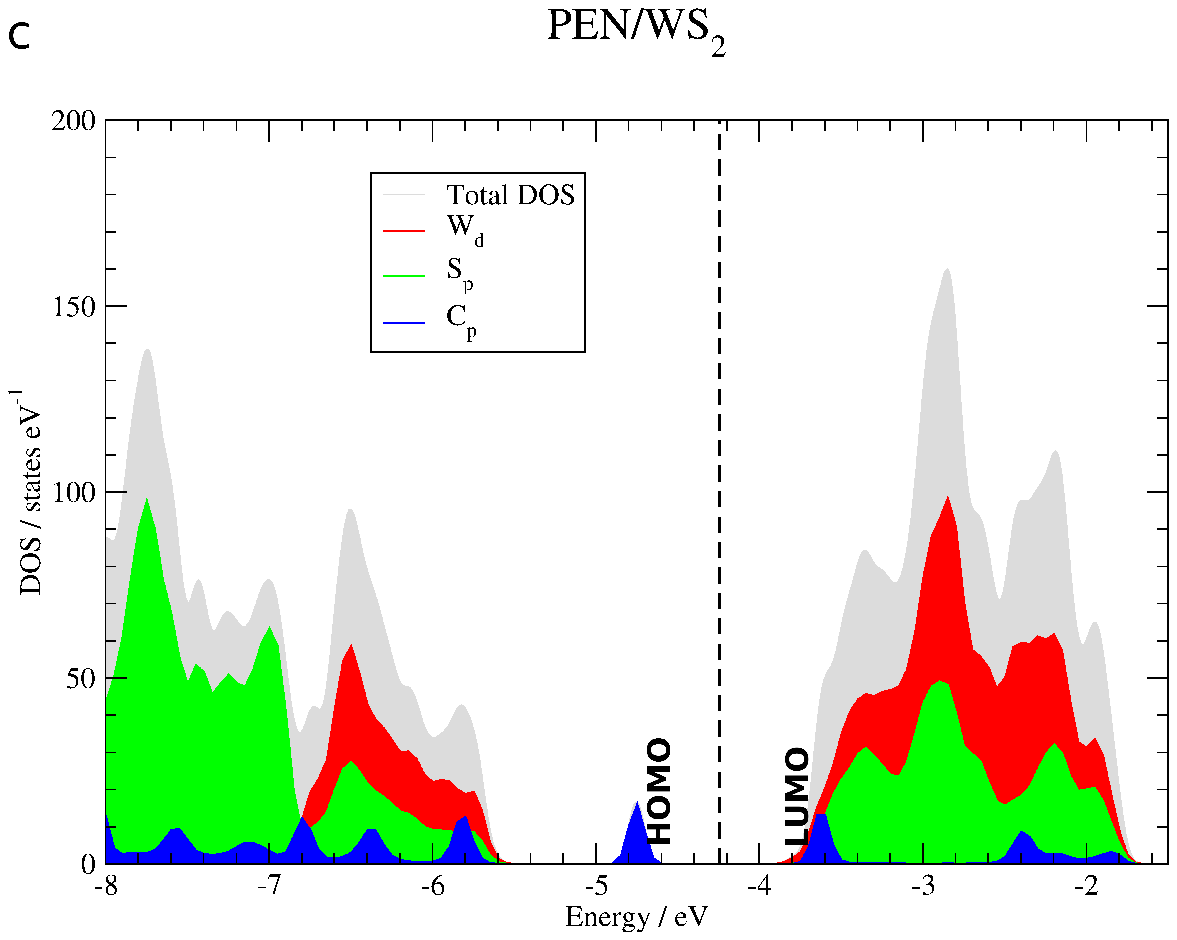}
	\includegraphics[scale=0.6]{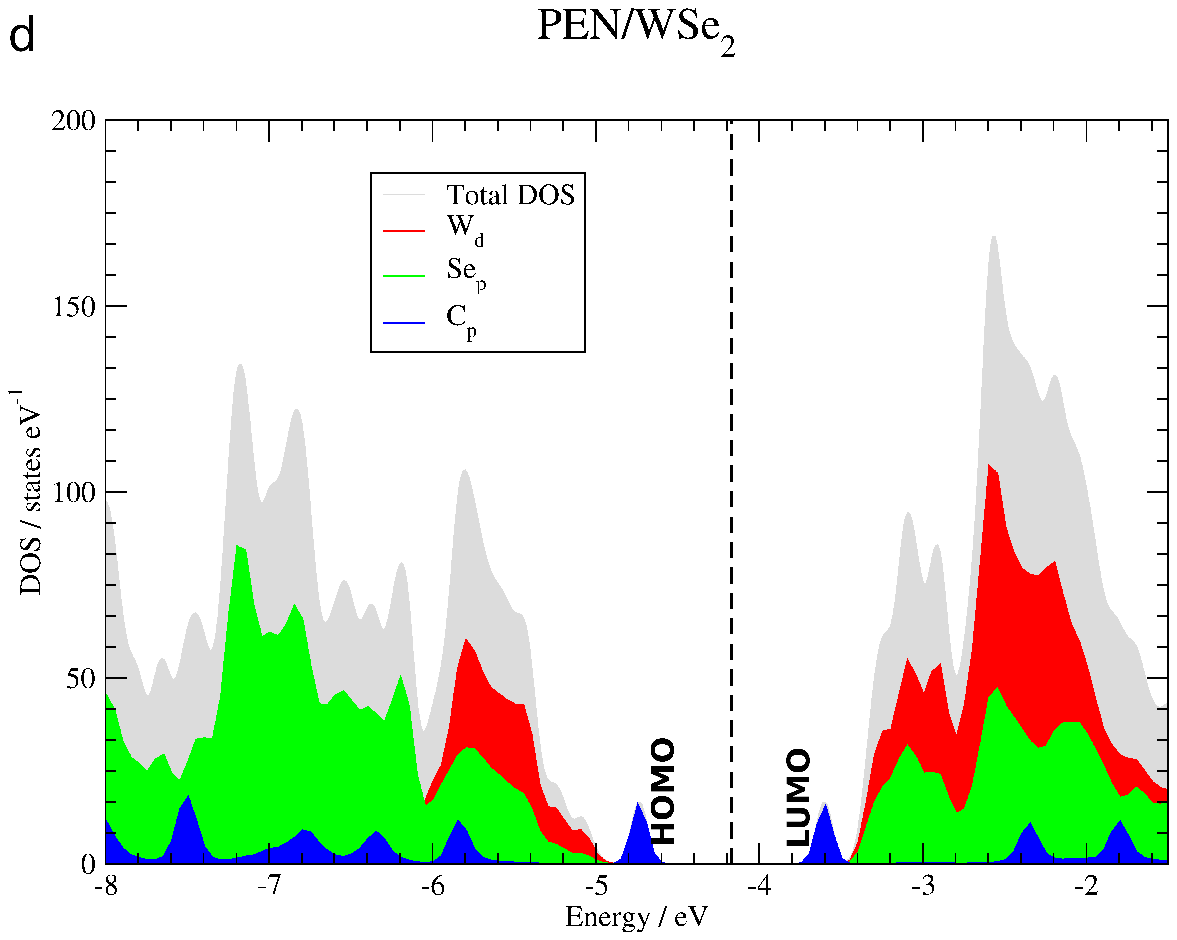}
	\caption{\label{fig-pdos}Total and partial density of states (DOS) of (a) PEN/MoS$_2$, (b) PEN/MoSe$_2$, (c) PEN/WS$_2$ and (d) PEN/WSe$_2$ heterostructures. The Fermi level of the heterostructure is indicated by a vertical dashed line. HOMO and LUMO of pentacene are highlighted in each plot.}
\end{figure*}

In addition to the structural properties, we also investigated the electronic properties of pentacene/TMD heterostructures, considering single-layer pentacene on top-Ch adsorption sites for all the four TMDs. Figure~\ref{fig-pdos} displays the density of states (DOS) of pentacene/TMD heterostructures (considering the most favourable adsorption site, namely, top-Ch), clearly showing that the highest occupied molecular orbital (HOMO) of pentacene is located within the band gap of the 2D TMDs, closer to the valence band maximum (VBM) of the selenide systems, MoSe$_2$ (Fig.~\ref{fig-pdos}(b)) and WSe$_2$(Fig.~\ref{fig-pdos}(d)) when compared to the sulfide systems, MoS$_2$ (Fig.~\ref{fig-pdos}(a)) and WS$_2$(Fig.~\ref{fig-pdos}(c)). This helps to explain why the interaction between pentacene and the Se-system is stronger than that of pentacene and S-systems. 
The lowest unoccupied molecular orbital (LUMO) of pentacene is located above the conduction band minimum (CBM) of MoS$_2$, MoSe$_2$ and WS$_2$, indicating that PEN/MoS$_2$, PEN/MoSe$_2$ and PEN/WS$_2$ form staggered type-II heterostructures; however, pentacene's LUMO has lower energy than the CBM of WSe$_2$ suggesting a type-I band alignment for PEN/WSe$_2$ heterostructure. By examining the DOS of the isolated molecule and isolated monolayer WSe$_2$ (see Fig.~S1 of Supplementary Material) we notice that the isolated systems have a type-II band alignment; however, molecule-molecule interactions present when the pentacene molecules are placed in a 7x4 supercell causes the HOMO and LUMO to shift by about 88~meV and 98~meV to lower energy, respectively, while molecule-substrate interaction causes an additional shift to lower energies of 209~meV for the HOMO and 194~meV for the LUMO (see Fig.~S2 of Supplementary Material), which leads to a transition from type-II to type-I alignment.  

We have not considered the effect of spin-orbit coupling (SOC) in our calculations here. We do not expect significant changes in the adsorption energies and geometries, but for the electronic properties, SOC will cause splits in the topmost valence bands as well as in the lowest conduction bands of the TMDs, in particular for WS$_2$ and WSe$_2$. We have computed the shift of the VBM and CBM of the TMDs (see Table S4 of the Supplementary Material), in order to infer if SOC would affect the type of band alignment observed here. For MoS$_2$, the shifts of VBM and CBM due to SOC are less than 0.1~eV, while pentacene HOMO is located about 1~eV above the VBM of MoS$_2$ and pentacene LUMO is located about 0.5~eV above the CBM of MoS$_2$, which indicates that SOC will not change the type-II band alignment observed here. Type-II band alignment is also expected to be preserved for both MoSe$_2$ and WS$_2$, since SOC causes shifts of the MoSe$_2$ and WS$_2$ VBM by about 0.23 and 0.17~eV to higher energies, respectively, which is not enough to place VBM above pentacene HOMO (which is located about 0.4~eV above the VBM of MoSe$_2$ and 0.8~eV above the VBM of WS$_2$). As expected, WSe$_2$ exhibits the largest shifts due to SOC: the VBM is shifted by about 0.33~eV to higher energies while the CBM is shifted by about 0.15~eV to lower energies. Since pentacene HOMO is located about 0.2~eV above WSe$_2$ VBM and pentacene LUMO is located about 0.2~eV below WSe$_2$ CBM, we expect that SOC will cause a larger hybridization between pentacene HOMO and WSe$_2$ VBM as well as between pentacene LUMO and WSe$_2$ CBM, facilitating charge transfer between these systems and potentially restoring the type-II band alignment. 

\begin{figure*}[!h]
	\centering
	\includegraphics[scale=0.25]{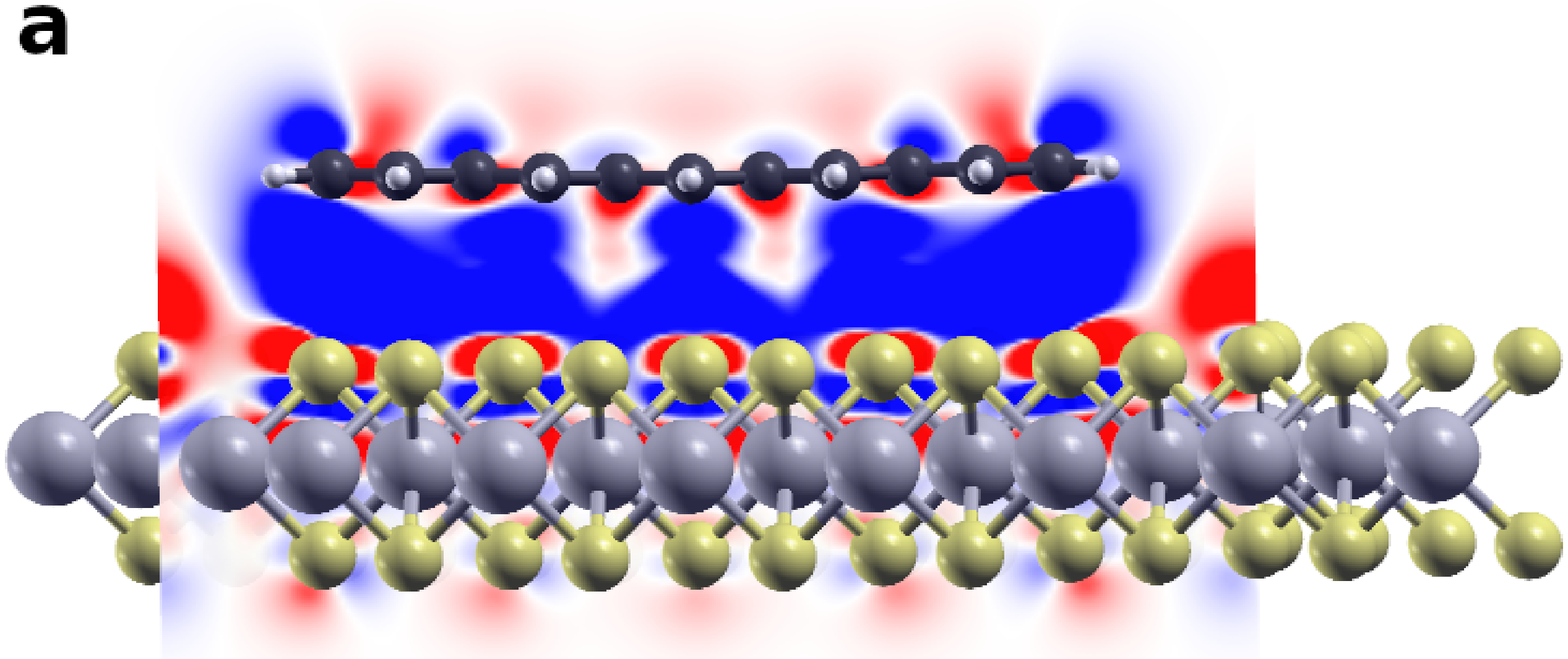}
	\includegraphics[scale=0.25]{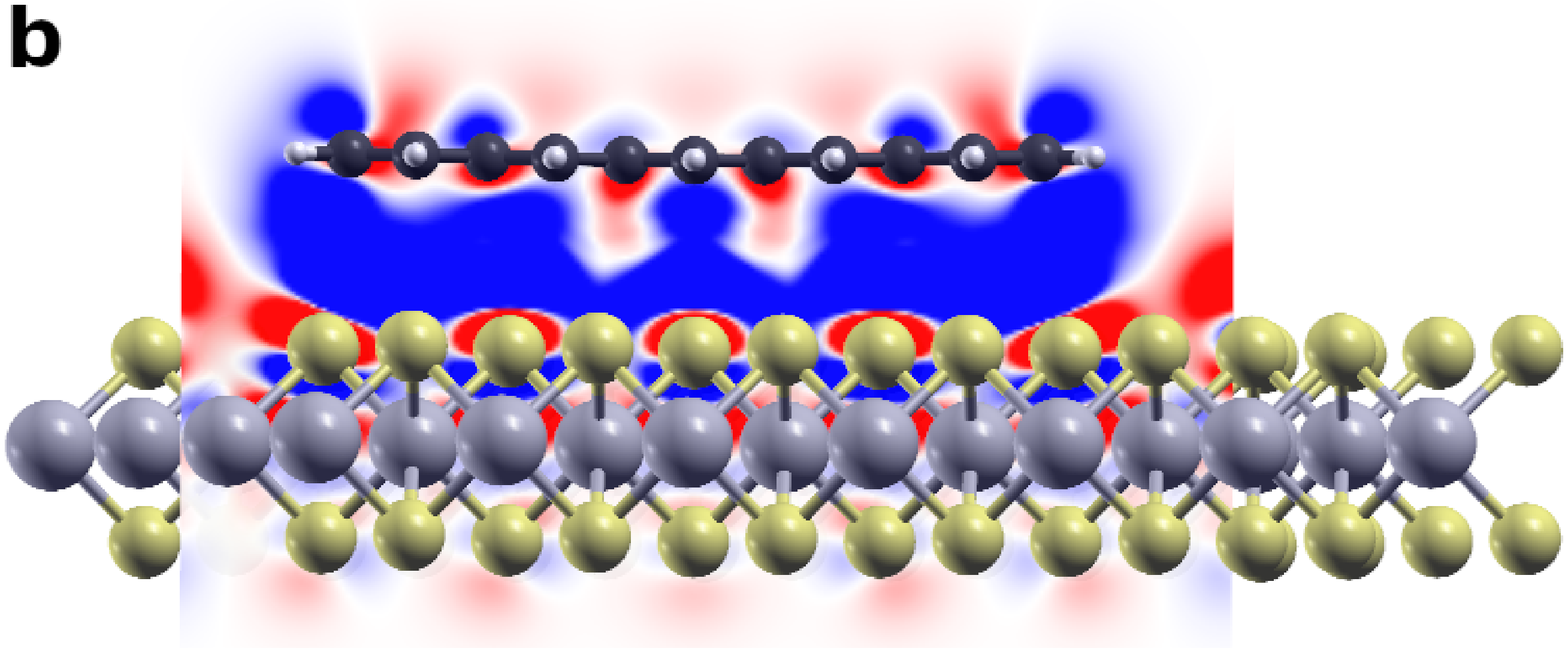}
	\includegraphics[scale=0.25]{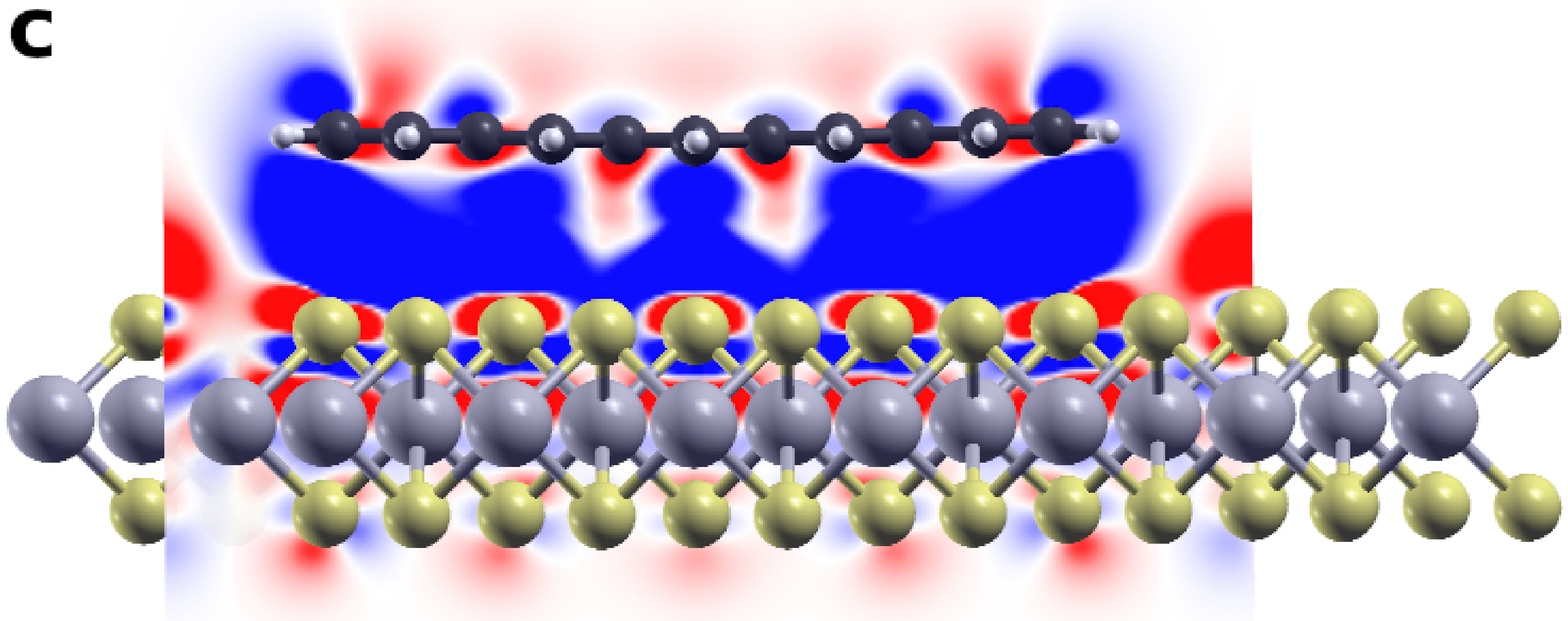}
	\includegraphics[scale=0.25]{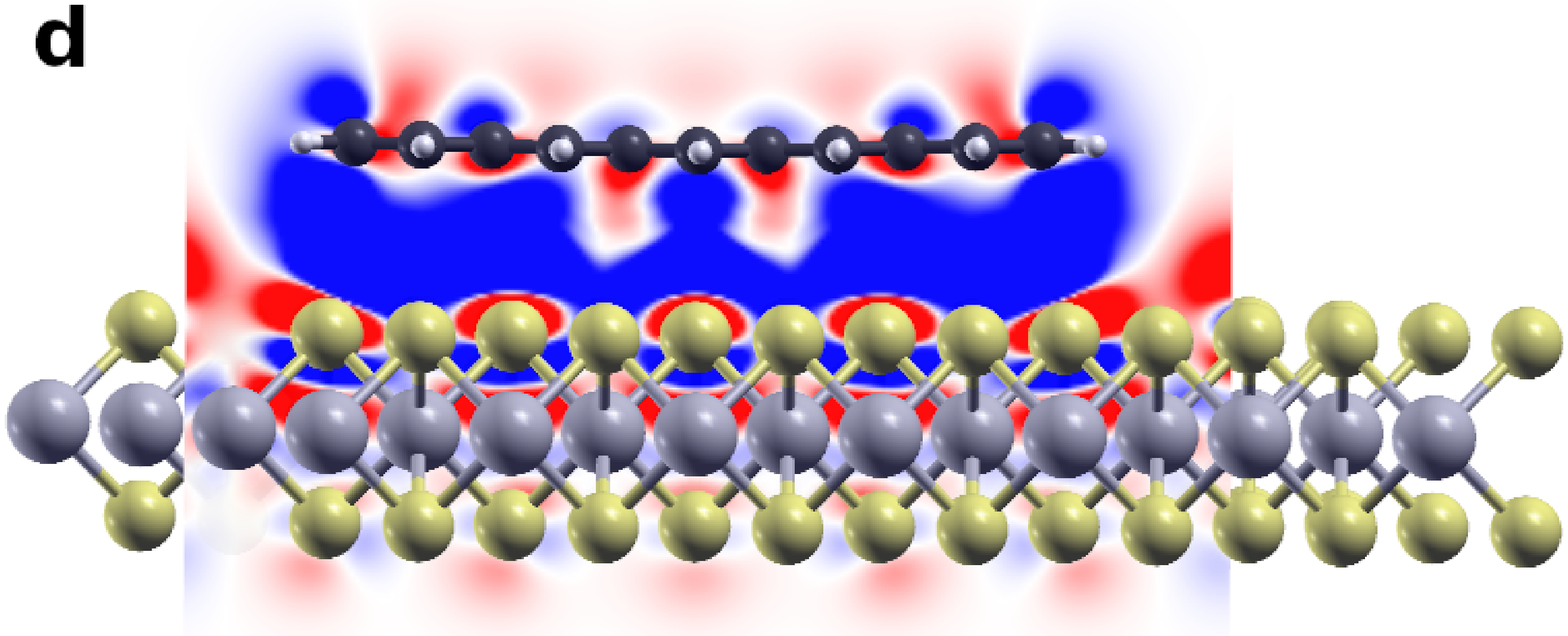}
	\caption{\label{fig-charge-2D}Charge density difference between the heterostructures ((a) PEN/MoS$_2$, (b) PEN/MoSe$_2$, (c) PEN/WS$_2$ and (d) PEN/WSe$_2$) and the isolated systems plotted on a plane perpendicular to both the pentacene molecule and the 2D TMD monolayer, cutting through the long axis of the molecule. Regions in blue and red represent depletion and accumulation of charge, respectively. Figure prepared using the XCrySDen software\cite{xcrysden}, with isovalues in the range of -0.00008 (blue) and +0.00008 (red) 
e/bohr$^3$.}
\end{figure*}

\begin{figure*}[!h]
	\centering
	\includegraphics[scale=0.3]{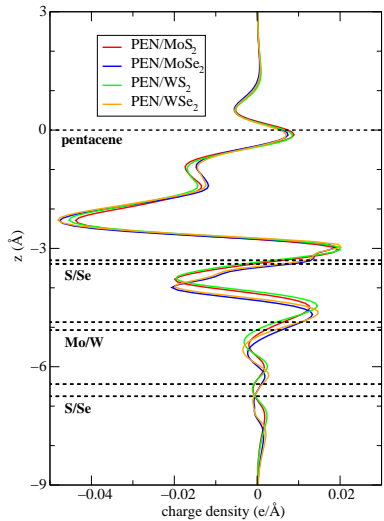}
	\caption{\label{fig-charge-integration}Charge density difference integrated along the horizontal direction of the 2D plots shown in Fig.~\ref{fig-charge-2D} plotted along the z direction. Horizontal dashed lines show the positions of the pentacene molecule, and the top/bottom chalcogen layer and the transition-metal layer of the TMD monolayer.}
\end{figure*}

Finally, we examine the charge transfer between pentacene molecules and the 2D TMD systems. Charge density difference between the heterostructure and isolated systems (Fig.~\ref{fig-charge-2D}), plotted on a plane cutting the long axis of the pentacene molecule and perpendicular to both molecule and TMD monolayer, shows characteristics of a Pauli repulsion pillow effect for all the systems: the overlap between the electronic clouds of the molecule and the TMD causes the charge to be pushed back into the TMD and around the edges of the molecules---red regions indicate accumulation of charge in the top chalcogen layer and around the edges of the molecules---leading to a depletion of charge (blue region) between the molecule and the 2D material. The plot of the charge density difference integrated along the horizontal direction, as displayed in Fig.~\ref{fig-charge-integration}, shows that 
the pentacene molecule gets polarized leading to an increased charge density on the side oriented towards the TMD, but there is no net charge transfer to or from the molecule.
The charge depletion between the molecule and the TMD is slightly larger in the Se systems when compared with the S systems, which is a result of the stronger interaction between pentacene and Se systems. The push-back of charge into the top chalcogen layer has the same amplitude in all four systems, as we can see in the region between $z=-3$ and $z=-4$~{\AA} (Fig.~\ref{fig-charge-integration}) where the top chalcogen layer is located. 
%The larger depletion of charge in the space between molecule and TMD observed in the Se systems also explains the smaller change in the work function caused by the adsorption of pentacene on MoSe$_2$ and WSe$_2$ when compared to MoS$_2$ and WS$_2$, as shown in Table~\ref{table-work-function}. This is because the push-back effect causes the electrons 
%
%The push-back effect of the charge density is also reflected in 
The reduction of the work function observed upon adsorption of the pentacene molecule, as seen in Table~\ref{table-work-function}, originates mainly from the HOMO level of the molecule lying higher than the valence band edge of the TMDs, cf. Fig.~\ref{fig-pdos}. 
This is in line with the observation that 
the change in the work function of the Se-systems is smaller than that observed for S-systems;
%, which is a result of the stronger interaction between pentacene and both MoSe$_2$ and WSe$_2$; this stronger interaction counteracts 
in addition, the push-back effect caused by the adsorption, which is similar in the S and Se compounds, contributes to the decrease of the work function.
%{\it The work functions calculated here for bare monolayer MoS$_2$, MoSe$_2$, WS$_2$ and WSe$_2$ (5.34, 4.46, 4.82, and 4.40 eV, respectivelly) are in good agreement with those obtained in previous DFT calculations reported in Ref.~\onlinecite{Kim-PRB103-085404} (5.23, 4.65, 4.85, and 4.38 eV, respectivelly).} 

\begin{table*}[!h]
	\caption{\label{table-work-function}Work function ($\phi$) of the TMDs and the PEN/TMD heterostructures.}  
\centering
\begin{tabular}{lccc}
\hline
	       &  $\phi$(TMD) (eV) & $\phi$(PEN/TMD) (eV) & $\Delta \phi = \phi({\rm TMD}) - \phi({\rm PEN/TMD}) $  (eV)\\
\hline
	MoS$_2$		&	5.34 & 4.42 & 0.92\\
	MoSe$_2$ 	&	4.46 & 4.21 & 0.25\\
	WS$_2$		& 	4.81 & 4.25 & 0.57\\
	WSe$_2$		&	4.40 & 4.17 & 0.23\\
\hline
\hline
\end{tabular}
\end{table*}

\section{Conclusions}

In summary, we have investigated the adsorption of single molecule pentacene on 2D monolayer TMDs (MoS$_2$, MoSe$_2$, WS$_2$ and WSe$_2$). Our results show that pentacene lies flat in all four TMDs, and interacts more strongly with MoSe$_2$ and WSe$_2$, which is reflected in a larger depletion of charge in the region between the molecule and the substrate. This also results in a smaller change in the work function caused by the adsorption of pentacene on MoSe$_2$ and WSe$_2$ when compared to MS$_2$ and WS$_2$, since the larger depletion of charge cancels out part of the push-back effect caused by the overlap between the electron densities of the molecule and the TMDs. 
Finally, we found that pentacene in a monolayer concentration forms type-II band alignment with MoS$_2$, MoSe$_2$ and WS$_2$, since pentacene HOMO has higher energy than the VBM of these systems, while the CBM of the TMDs have lower energy than pentacene LUMO. For PEN/WSe$_2$ we observed a type-I band alignment, since molecule-molecule and molecule-substrate interactions shift pentacene LUMO to lower energies, placing it below the CBM of WSe$_2$. The discussion of the interaction between pentacene molecules and monolayer TMDs is valuable since such heterostructures show promising electronic properties for application in flexible, lightweight and thin photovoltaics. 

\section{Author contributions}
JM and PK conceived the project. EB and JM performed the calculations. JM supervised the project. 
All authors analysed and discussed the results and contributed to the writing of the manuscript. 
The authors declare no competing financial interest.
All authors have given approval to the final version of the manuscript.

\section{Acknowledgement}
This work was partially funded by the Deutsche Forschungsgemeinschaft (DFG, German Research Foundation), Project 406901005, and used the Cirrus UK National Tier-2 HPC Service at EPCC (http://www.cirrus.ac.uk) funded by the University of Edinburgh and EPSRC (EP/P020267/1). 

\section{Data availability statement}
The data that support the findings of this study are openly available at Keele Data Repository: \url{https://doi.org/10.21252/j6dg-m071}. 

\bibliography{references.bib}
\bibliographystyle{ieeetr}
\end{document}